\documentclass[journal,twoside,web]{ieeecolor}
\usepackage{generic}
\usepackage{cite}
\usepackage{amsmath,amssymb,amsfonts}
\usepackage{graphicx}
\usepackage{textcomp}
\usepackage{csquotes}
\usepackage{acronym}
\usepackage{subfigure}
\usepackage{xcolor}
\usepackage[colorlinks]{hyperref}
\hypersetup{
    linkcolor=blue,
    filecolor=magenta,      
    urlcolor=teal,
    citecolor=magenta
    }
\usepackage{dblfloatfix}
\usepackage{algorithm}
\usepackage{algpseudocode}

\usepackage{etoolbox}
\makeatletter
\@ifundefined{color@begingroup}%
{\let\color@begingroup\relax
        \let\color@endgroup\relax}{}%
\def\fix@ieeecolor@hbox#1{%
        \hbox{\color@begingroup#1\color@endgroup}}
\patchcmd\@makecaption{\hbox}{\fix@ieeecolor@hbox}{}{\FAILED}
\patchcmd\@makecaption{\hbox}{\fix@ieeecolor@hbox}{}{\FAILED}

\acrodef{SVM}{support vector machine}
\acrodef{PCG}{phonocardiogram}
\acrodef{RF}{random forest}
\acrodef{ANN}{artificial neural network}
\acrodef{DNN}{deep neural network}
\acrodef{CNN}{convolutional neural network}
\acrodef{RNN}{recurrent neural network}
\acrodef{LSTM}{long short-term memory}
\acrodef{MFCCs}{Mel-frequency cepstral coefficients}
\acrodef{GRU}{gated recurrent unit}
\acrodef{CAD}{computer-assisted decision}

\definecolor{myClr}{rgb}{0.6350, 0.0780, 0.1840}

\def\BibTeX{{\rm B\kern-.05em{\sc i\kern-.025em b}\kern-.08em
    T\kern-.1667em\lower.7ex\hbox{E}\kern-.125emX}}
\begin{document}

\title{Beyond Heart Murmur Detection: Automatic Murmur Grading from Phonocardiogram}

\author{Andoni Elola, Elisabete Aramendi, \IEEEmembership{Member, IEEE}, Jorge Oliveira, Francesco Renna, \IEEEmembership{Senior Member, IEEE}, Miguel T. Coimbra, \IEEEmembership{Senior Member, IEEE}, Matthew A. Reyna, Reza Sameni, \IEEEmembership{Senior Member, IEEE}, Gari D. Clifford, \IEEEmembership{Fellow, IEEE}, and  Ali~Bahrami~Rad
\thanks{AE and EA are  supported by the Spanish Ministerio de Ciencia,
Innovaci\'on y Universidades under Grant RTI2018-101475-BI00, jointly
with the Fondo Europeo de Desarrollo Regional (FEDER), by the
Basque Government under Grant IT1717-22 and by the University of the Basque Country (UPV/EHU) under Grant
COLAB20/01. The work of FR and MC is financed by National Funds through the Portuguese funding agency, FCT - Funda\c{c}\~{a}o para a Ci\^{e}ncia e a Tecnologia, within project UIDB/50014/2020. GC, MR and ABR received grant funding from the National Institute of Biomedical Imaging and Bioengineering (NIBIB) under NIH grant R01EB030362, and the National Center for Advancing Translational Sciences of the National Institutes of Health under Award Number UL1TR002378, as well as donations from Alivecor Inc. and Mathworks Ltd.}
% GC and MR are partially supported by the National Institute of General Medical Sciences (NIGMS) and the National Institute of Biomedical Imaging and Bioengineering (NIBIB) under NIH grant numbers 2R01GM104987-09 and R01EB030362 respectively, the National Center for Advancing Translational Sciences of the National Institutes of Health under Award Number UL1TR002378.
\thanks{Andoni Elola is with the Department of Electronic Technology, University of the Basque Country, Eibar, Gipuzkoa, Spain and also with Department of Biomedical Informatics, School of Medicine, Emory University, Atlanta, GA, USA (e-mail: andoni.elola@ehu.eus). }
\thanks{Elisabete Aramendi is with the University of the Basque Country, Bilbao, Spain.}
\thanks{Jorge Oliveira is with Universidade Portucalense Infante D. Henrique, Porto, Portugal (e-mail: jorgefisicomat@gmail.com).}
\thanks{Francesco Renna and Miguel T. Coimbra are with INESC TEC, Faculdade de Ciências da Universidade do Porto, Porto, Portugal (e-mail: \{francesco.renna, mcoimbra\}@fc.up.pt).}\thanks{Matthew Reyna, Reza Sameni, Gari D. Clifford, and Ali Bahrami Rad are with the Department of Biomedical Informatics, School of Medicine, Emory University, Atlanta, GA, USA. Gari D. Clifford is also with the Department of Biomedical Engineering, Georgia Institute of Technology and Emory University, Atlanta, GA, USA}}

\maketitle

\begin{abstract}
%\commentAE{Reza, can you write this section please?}

\textit{Objective:} Murmurs are abnormal heart sounds, identified by experts through cardiac auscultation. The \textit{murmur grade}, a quantitative measure of the murmur intensity, is strongly correlated with the patient's clinical condition. This work aims to estimate each patient's murmur grade (i.e., absent, soft, loud) from multiple auscultation location phonocardiograms (PCGs) of a large population of pediatric patients from a low-resource rural area. \textit{Methods:} The Mel spectrogram representation of each PCG recording is given to an ensemble of 15 convolutional residual neural networks with channel-wise attention mechanisms to classify each PCG recording. The final murmur grade for each patient is derived based on the proposed decision rule and considering all estimated labels for available recordings. The proposed method is cross-validated on a dataset consisting of 3456 PCG recordings from 1007 patients using a stratified ten-fold cross-validation. Additionally, the method was tested on a hidden test set comprised of 1538 PCG recordings from 442 patients. \textit{Results:} The overall cross-validation performances for patient-level murmur gradings are 86.3\% and 81.6\% in terms of the unweighted average of sensitivities and F1-scores, respectively. The sensitivities (and F1-scores) for absent, soft, and loud murmurs are 90.7\% (93.6\%), 75.8\% (66.8\%), and 92.3\% (84.2\%), respectively. On the test set, the algorithm achieves an unweighted average of sensitivities of 80.4\% and an F1-score of 75.8\%. \textit{Conclusions:} This study provides a potential approach for algorithmic pre-screening in low-resource settings with relatively high expert screening costs.  \textit{Significance:} The proposed method represents a significant step beyond detection of murmurs, providing characterization of intensity, which may provide an enhanced classification of clinical outcomes. 
\end{abstract}

\begin{IEEEkeywords}
Murmur, Murmur grading, Phonocardiogram (PCG), Mel Spectrogram, Convolutional Neural Networks, Uncertainty
\end{IEEEkeywords}

\section{Introduction}

% \begin{itemize}
%     \item Impact of cardiac diseases on society, especially focusing on pediatric cardiac diseases and underprivileged countries scenarios
%     \item Digital auscultation, as a non-invasive screening tool, can play a fundamental role in reducing neonatal death rate in these scenarios 
%     \item Computer assisted decision systems can enhance the impact of cardiac auscultation: heart sound classification
%     \item Binary vs. murmur grading: advantages from the clinical point of view.
%     \item Majority of classification algorithm in the literature are binary...
%     \item Our contributions
%     \item Paper organization and notation (if needed)
% \end{itemize}

%\commentFR{work in progress...}

\IEEEPARstart{C}{ardiovascular} diseases are the leading cause of death worldwide, accounting for approximately $31\%$ of all global deaths \cite{WHO:2017}. In high-income countries, coronary artery diseases are more prevalent; on the other hand, congenital and heart diseases have higher prevalence in low and middle income countries, in which the healthcare system is overwhelmed and patient prescreening and triage is
inevitable. Low and middle income countries also face challenges in diagnosing and treating both congenital and acquired heart conditions. This is mainly due to the lack of cardiologist specialists in vast areas, which are under-resourced and have limited access to health services \cite{Carvalho12}. In these settings, the majority of patients are never visited by an expert cardiologist.

Digital heart sound auscultation through the \ac{PCG} allows a non-invasive assessment of the mechanical activity of the heart, thus providing valuable early information regarding congenital and acquired diseases in children. In addition, digital cardiac auscultation, due to its low cost and simplicity, can be carried out in point-of-care scenarios, without requiring advanced training for heart sound collection. On the other hand, the interpretation of auscultation sounds requires intensive, prolonged training \cite{Dwivedi18, Mangione01}. Moreover, there are also significant differences between the standards for PCG diagnosis across different healthcare settings and countries. As a result, PCG-based
diagnosis of heart abnormalities remains highly subjective. 
These factors have recently spurred a renewed interest in developing devices powered by machine learning algorithms for automatic heart sound analysis that can help nurse practitioners and
junior medical doctors with the triage of the cases that are suspicious of serious heart abnormalities, which can be transformative and life-saving globally and more prominently in low-resourced areas.%to be possibly integrated into computer aided decision-support systems for cardiovascular disease screening.

To this end, several solutions have been recently presented that automatically classify heart sounds. However, most of these methods focus on binary classification, thus providing information only regarding normal vs.\ abnormal heart sounds, or the presence vs.\ absence of heart murmurs. On the other hand, the existing research efforts, which have attempted to provide richer descriptions of heart sounds, are limited to a small set of specific heart sound-inferred  diseases. For a general overview of recent solutions in automatic heart sound classification, the reader is referred to \cite{Clifford17, Dwivedi18, chen2021deep} and references therein.

In contrast with the majority of the current solutions for automatic heart sound classification, clinical practice in cardiac disease screening via auscultation usually consists of providing a detailed characterization of the possibly present murmurs by considering different aspects, including timing, shape, pitch, and quality of the sounds. In particular, the Levine scale is commonly used by clinicians to evaluate the severity of systolic murmurs~\cite{Levine:1933}. This scale represents a numeric score ranging from 1 to 6 (from I/VI to VI/VI, using the standard clinical notation for grading), which is associated with the intensity and loudness of the murmur, as well as the locations from which the murmur can be detected during the auscultation process or during a physical exam. The information carried by the analysis of the grading of murmurs with the Levine scale is extremely important to detect heart defects, as louder murmurs (grade $\geq$ III/VI) are more likely to be associated with cardiac defects \cite{Keren2005}. However, murmur intensity is typically assessed by comparing one murmur to another, for which no commonly accepted gold standard has been established. Therefore, the evaluation of murmur grading can be affected by ambient noise, patient anatomy (e.g., thicknesses of the patients' chest walls), and subjective judgments of the clinician \cite{Keren2005}. These factors strongly motivate the development of automatic murmur grading tools capable of providing more precise, robust, consistent, and objective outcomes.

The aim of this study is to develop a novel automated algorithm to characterize a patient's murmur severity grade in three categories: \textit{absent} (no murmur detected), \textit{soft} (murmur with Levine's grade I and II) and \textit{loud} (murmur with Levine's grade III and above). This is an attempt to provide automatic analysis of heart sounds related to various pathologies and deviates from previous multi-class PCG classification solutions, which attempted to directly link heart sounds with specific pathologies (described in Section~\ref{par:sota}). The proposed classification approach has the advantage of being suitable for the implementation in a computer aided decision-support system for auscultation-based cardiovascular screening, given that it provides a clinically-based explainable characterization of murmurs. The proposed approach is evaluated over a large dataset of heart sounds collected in a real-world auscultation setting, thus allowing the consideration of the effect of various sources of noise in the automatic analysis of heart sounds.

In particular, the contributions of this research include: A) defining a novel heart sound multi-class classification problem using the definition of murmur grading based on the clinically accepted Levine scale, B) developing a deep learning solution for automatic murmur grading based on a residual convolutional neural network (CNN) architecture and channel attention mechanisms, and C) the evaluation of the performance of the proposed solution for automatic murmur grading on a large dataset that contains PCG recordings from multiple auscultation locations, in real-world scenario with frequent noise and disturbances.

%The remainder of this paper is organized as follows: Section~\ref{par:sota} describes prior studies on PCG processing. In Section~\ref{par:database} the description of the utilized dataset is provided. In Sections~\ref{par:methods} and \ref{par:results} the methodology and the results are presented. Section~\ref{par:discussion} is dedicated to the discussion about the results, limitations and future work. Concluding remarks are presented in Section~\ref{par:conclusions}.

\section{Prior studies}
\label{par:sota}

We review the state-of-the-art on binary and multi-class heart sound classification.

\subsection{Binary Heart Sound Classification}
To date, numerous algorithms have been proposed to discriminate between normal and abnormal heart sounds, or to detect the presence of murmurs in \ac{PCG} signals. Many approaches focus on designing \emph{ad hoc} features extracted from the data for the \ac{PCG} and machine learning classifiers \cite{Ari10,Bhatikar05,Zabihi16}, while other approaches are based on deep learning solutions \cite{chen2021deep}. These approaches do not require the computation of \emph{ad hoc} discriminative features, since the classifiers can be directly applied to the \ac{PCG} in the time domain \cite{Zhang16,Ryu16,Xiao20,Thomae:2016,kiranyaz2020real,potes2016ensemble,Latif:2018} or to some time-frequency representation of the \ac{PCG} \cite{maknickas2017recognition,Nilanon16,Rubin:2017,Zhang19}, thus implementing an end-to-end system. For both kinds of solutions, \ac{PCG} classification approaches are often preceded by a segmentation step, which identifies the S1 sounds, the systole interval, the S2 sounds and the diastole interval in each heartbeat. This step is useful to identify the position of murmurs, potentially leading to the extraction of more significant features. On the other hand, the presence of murmurs and the auscultation environmental conditions, especially in point-of-care scenarios, often make the segmentation task challenging \emph{per se}.

Some of the prior studies included the murmur detection in the segmentation algorithm pipeline \cite{pedrosa2014automatic, varghees2017effective}, while other studies focused solely on the detection of the presence of murmurs \cite{das2022deep,oliveira2021multi,alam2018murmur}. %In \cite{alam2018murmur}, a recurrent neural network (RNN) and a CNN were used in parallel to discriminate between {normal} heart sounds and heart sounds recordings with {murmurs}. It reported an F1-score of 98\%. In \cite{das2022deep} a deep neural network was fed using 64-dimensional acoustic features for murmur detection. The reported F1-score was 98.3\%. 

\subsection{Multi-class Heart Sound Classification}
Several attempts have been made in the literature to provide a more complete characterization of heart sounds beyond simple anomaly detection. 
%\commentAE{Segmenting the PCG signal in out-of-hospital environments is more challenging, and even more when murmurs are present. We should highlight this too.}\commentFR{I agree}
As with the case of binary classification, multi-class approaches can be also categorized into methods requiring manual design of discriminative features vs.\ more recent deep learning approaches. In addition, some research efforts have focused on characterizing different aspects of heart sound recordings, such as specific heart valve diseases \cite{Maglogiannis09,Safara13,Zheng15,Dong19}. These studies attempted to associate the presence of murmurs with different pathologies such as aortic stenosis, aortic regurgitation, mitral stenosis, etc.

Some studies performed a murmur detection task within a multi-class classification framework. For instance, the authors of \cite{chorba2021deep} attempted to automatically detect \textit{inadequate signals} besides presence/absence of murmurs, thus obtaining a sensitivity of 76.3\% and a specificity of 91.4\% for murmur detection. In \cite{Ahlstrom06}, 36 patients were classified according to the presence of physiological murmurs, mitral insufficiency or aortic stenosis. In \cite{raza2019heartbeat}, an attempt to discriminate between \textit{normal}, \textit{murmur} and \textit{extra-systole} sounds was made. In \cite{Vepa09}, different classifiers were analyzed to discriminate between \textit{normal}, \textit{systolic murmurs} and \textit{diastolic murmurs} sounds. 

More recently, many teams developed algorithms to discriminate between \textit{murmurs present}, \textit{absence of murmurs} and \textit{unsure} from multi-location PCGs for the George B. Moody Physionet Challenge 2022. Teams were evaluated in terms of a custom weighted accuracy metric, i.e. the teams with the highest scores on the hidden test set were the winners of the challenge. Details can be found in \cite{reyna2022heart}. The top three algorithms were \cite{lu2022lightweight,mcdonald2022detection,xu2022hierarchical}. In \cite{lu2022lightweight}, the mel-spectrogram along with some wide features (including demographic data, zero-crossing rate or spectral bandwidth) were used to feed a CNN with two branches. In \cite{mcdonald2022detection}, a hidden semi-Markov models and recurrent neural networks were used for murmur detection and robust PCG segmentation. Finally, \cite{xu2022hierarchical} proposed a hierarchical multi-scale CNN, spectrograms were calculated using different scales and they were combined in a single CNN.

However, to the best of our knowledge, there are no existing research that provide algorithms for fine murmur characterization according to clinically accepted grading criteria.

\section{The PCG Dataset}
\label{par:database}
%\subsection{Data description}
The dataset used for this study is a subset of a larger dataset, namely the CirCor DigiScope PCG dataset~\cite{CirCorPhysioNet,oliveira2021circor}. The dataset is a collection of heart sound signals collected over two mass screening campaigns in the state of Paraiba, Brazil, from July to August 2014 and from June to July 2015. The main goal of gathering this dataset was to investigate and categorize cardiac diseases in a pediatric and pregnant population. Contrary to other PCG datasets~\cite{Clifford17,clifford2016classification,EPHNOGRAM2021,kazemnejad2021open}, which typically consist of a single recording from a single precordial location for each subject, the CirCor dataset consists of multiple PCG recordings from multiple auscultation locations. For most patients in the dataset, the PCGs were recorded from four prominent auscultation locations: aortic valve, pulmonary valve, tricuspid valve, and mitral valve. However, some patients have recordings from fewer than four locations; on the other hand, few patients have multiple recordings per location. The recordings were collected sequentially (not simultaneously) from different locations. The number of recordings, their location, and their duration may vary between patients.

The entire dataset consists of 5272 PCG recordings from 1568 patients. The average age ($\pm$ standard deviation) of the participants is 6.1($\pm$4.3) years, ranging from 0 to 21 years. The PCGs were recorded with a sampling rate of 4000\,Hz, using the DigiScope Collector technology embedded in the Littmann 3200 stethoscope \cite{gomes2015proposal}. The minimum and the maximum recording lengths are 4.8\,s and 80.4\,s, respectively. The mean (standard deviation) heart rate is 102 ($\pm$20) beats per minute (bpm), ranging between 47 and 193\,bpm. A detailed description of the dataset can be found in \cite{CirCorPhysioNet} and \cite{oliveira2021circor}. %For this study, the utilized subset of the data corresponds to the training and validation sets of the George B. Moody PhysioNet Challenge 2022\footnote{The George B. Moody PhysioNet Challenge 2022: Heart Murmur Detection from Phonocardiogram Recordings, \url{https://moody-challenge.physionet.org/2022/}}. %However, to be in line with the aforementioned challenge and not to disclose any information about the test set except the statistics mentioned above, only a part of the data were used in this work.

The dataset is extensively annotated with detailed murmur characteristics. The annotations indicate the presence or absence of a murmur for each patient and provide a complete description of a murmur event, such as its location, most audible location, type, timing, shape, pitch, quality, and intensity grade. The expert annotator also labeled each record as murmur present, absent, or unknown (for low-quality records). 

\begin{figure*}[t]
    \centering
\subfigure[Absent]{\includegraphics[scale=1]{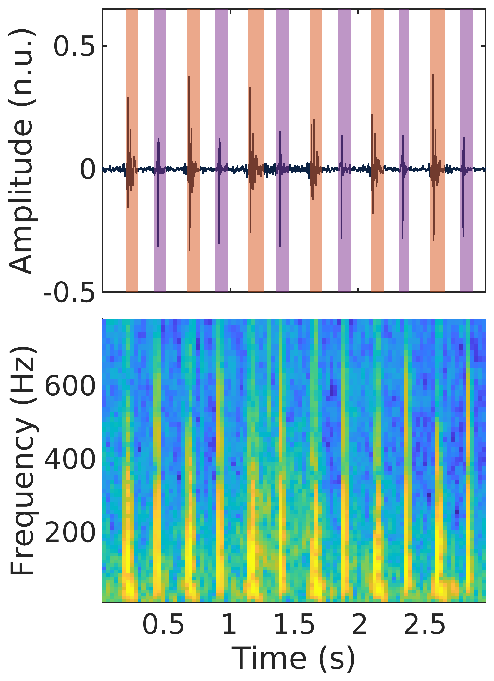}\label{fig:data_examples_absent}}
\subfigure[Soft]{\includegraphics[scale=1]{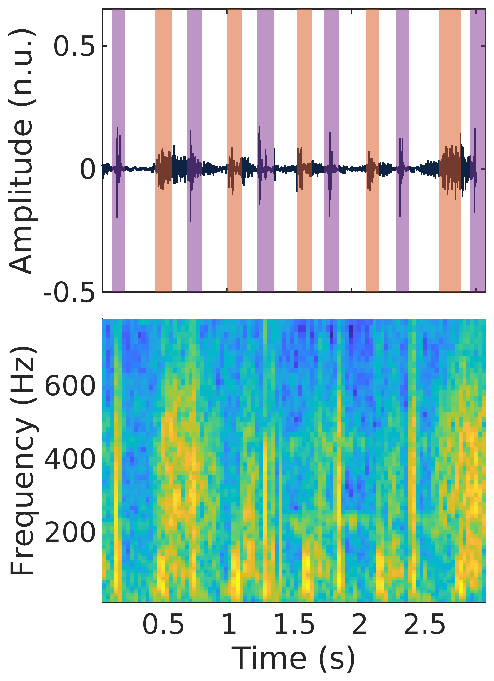}\label{fig:data_examples_soft}}
\subfigure[Loud]{\includegraphics[scale=1]{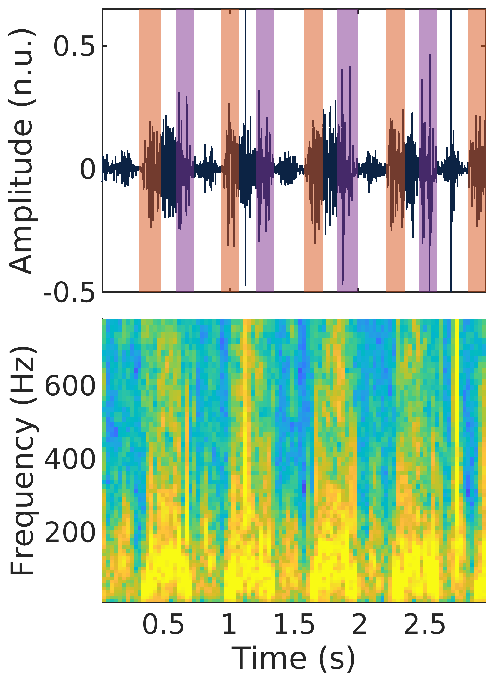}\label{fig:data_examples_loud}}
\caption{PCG examples in time domain and their corresponding mel spectrograms used as inputs of the deep neural network. In the time domain, S1 and S2 are highlighted in orange and purple, respectively, and n.u. indicates normalized units.}
\label{fig:data_examples}
\end{figure*}

The murmur grading was annotated by an expert clinician via listening to the audio recordings and visual inspection of the waveforms. Murmur gradings were grouped in the following three categories:
\begin{itemize}
    \item \textbf{Absent:} This label indicates that no murmurs are present in any auscultation location.
    \item \textbf{Soft:} This label corresponds to murmur grades I and II on the Levine scale \cite{Levine:1933}. In addition, by convention, if not all the locations were available and murmurs were present, it was considered that the patient had \textit{soft} murmurs.
    \item \textbf{Loud:} This label corresponds to murmur grades levels III or above, on the Levine scale.
\end{itemize}

For the annotation of the dataset, the Levine scale was slightly modified due to the following factors. First, grades above III require physical examination, but the annotator could only have access to the digital recordings. Therefore, grades above III were grouped as \textit{loud}. This was defined as \textit{more than soft} in other medical studies \cite{rishniw2018murmur,caivano2018murmur}. Second, by definition, grades above III correspond to murmurs that are audible in the main four locations; but in our dataset all four locations are not always available (missing data). In these cases, the patient was labeled as \textit{soft}. Lastly, grades I and II were grouped as \textit{soft} and were not further distinguished, as proposed in \cite{ljungvall2014murmur}.

A murmur grade was associated to each patient, and the annotator also labeled the location(s) in which the murmur could be heard. Moreover, in the dataset, most murmurs are present during the systolic phase~\cite{oliveira2021circor}. There is only one
patient with only diastolic murmurs, so due to the lack of
samples, we solely analyzed systolic murmurs in this study. The patient that only had diastolic murmurs was labeled as \textit{absent} and it was within the training set. %There is only one patient with only diastolic murmurs. Due to the lack of samples, we solely analyzed systolic murmurs in this study.

From the original dataset those patients labeled as \textit{unknown} (low-quality records) were discarded, which constituted less than 8\% of the data. In this study the training and validation sets of the George B. Moody Physionet Challenge 2022 data were considered to design the method\,\cite{reyna2022heart}. For this subset, after removing recordings with unknown cases of murmurs, 3456 recordings from 1007 patients were available, 492 male and 515 female. From these 1007 patients, 802 had no murmurs (\textit{absent}), 153 had \textit{soft} murmurs and 52 had \textit{loud} murmurs. Each patient contained a mean ($\pm$standard deviation) of 3.4 ($\pm$1.0) recordings, and the duration of the recordings ranged between 5.2-64.5\,s, with a mean ($\pm$standard deviation) of 22.8 ($\pm$7.3)\,s.

In addition, the hidden test data of the George B. Moody Physionet Challenge 2022 was used to validate the algorithm. After removing the unknown cases, the hidden test data contained a total of 1538 PCG recordings from 442 patients (241 male and 201 female), 233 without any murmur, 50 with \textit{soft} murmurs and 21 with \textit{loud} murmurs. Each patient contained a mean ($\pm$standard deviation) of 3.5 ($\pm1$) recordings. The duration of the recordings ranged between 4.8 and 80.4\,s, with a mean ($\pm$standard deviation) of 23.2 ($\pm$7.5).

\section{Methods}

\label{par:methods}
The goal of this study is to estimate the murmur's severity grade (\textit{absent}, \textit{soft}, or \textit{loud}) for each patient. As an intermediate step, we also evaluate the murmur severity (grade) for each PCG recording (murmur locations were also annotated). %This intermediate task can be considered the secondary goal of this study.  

%To do so, the recordings obtained from the available auscultation spots are analyzed jointly. 

In order to classify each recording, 2-D deep convolutional neural networks (CNN)s were used. This section includes a description of the method, a simple data visualization, data preprocessing steps, an elaboration of the classification model, a final decision rule for murmur grading, and the evaluation metrics.

Figure \ref{fig:data_examples} demonstrates typical PCG examples of different murmur grades in time and time-frequency (Mel spectrogram; cf. Section~\ref{sec:preprocessing}) domains that the algorithm of this study tries to discriminate.

\subsection{Data Visualization}
Before discussing the proposed method for classifying the PCG recordings based on the CNNs, a simple quantitative analysis among different murmur grades was conducted for visualization purposes and a better understanding of the nature of the problem.  

\begin{figure*}[tb]
    \centering
\subfigure[]{\includegraphics[scale=1]{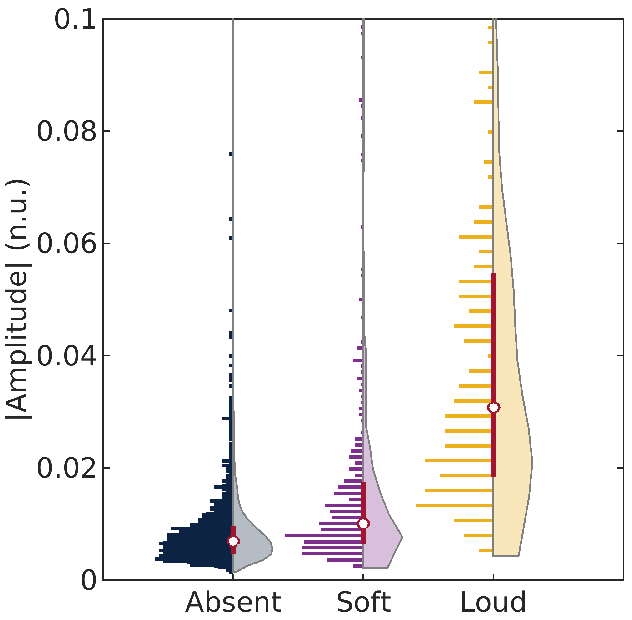}\label{fig:quantitative_analysis_hist}}
\subfigure[]{\includegraphics[scale=1]{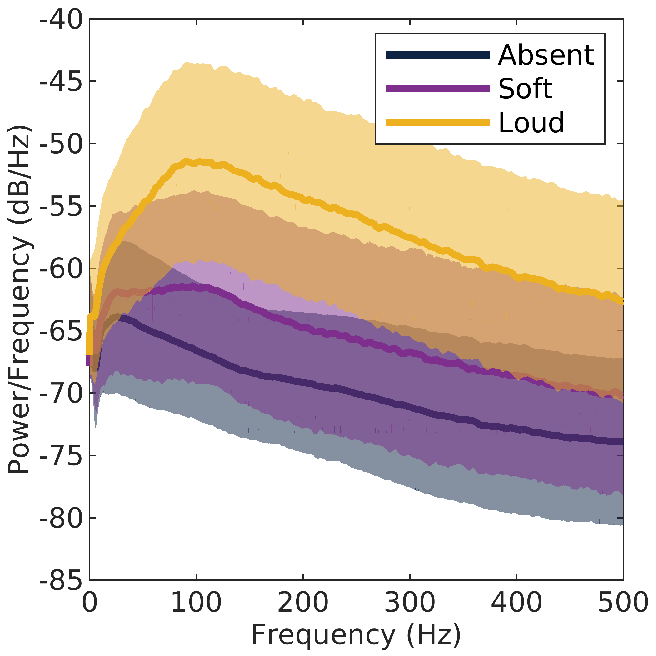}\label{fig:quantitative_analysis_dist}}
\caption{Results of the quantitative analysis: (a) mean absolute amplitude of the systolic phase per class, and (b) mean$\pm$standard deviation (shades) power spectra of classes. In the figure n.u. indicates normalized units.}
\label{fig:quantitative_analysis}
\end{figure*}

Using the PCG segmentation annotations provided in the dataset, we compared the mean absolute amplitudes of the systolic phase for different murmur grades across all recordings. The corresponding distributions are shown in Figure~\ref{fig:quantitative_analysis_hist}. Accordingly, \textit{absent} and \textit{loud} classes are well separated, while identifying \textit{soft} murmurs is more challenging, as the distributions of \textit{absent} and \textit{soft} murmurs have significant overlap. We also computed the mean and standard deviation of power spectra across all recordings for each class, and the result is shown in Figure~\ref{fig:quantitative_analysis_dist}. To calculate the spectrum of each recording, the mean power spectrum among systolic phases was computed using the periodogram. Again, by visual inspection, discriminating between \textit{soft} murmurs and heart sounds without murmurs appears to be challenging.

\subsection{Preprocessing}
\label{sec:preprocessing}

Each PCG recording was split using a sliding window of 3\,s without overlapping, and a 2-D representation of each segment was obtained by using the logarithmic Mel spectrogram \cite{rabiner2010theory}. The spectrogram was calculated using parameters used in previous studies \cite{Zabihi16,Rubin:2017}: Hamming windows of 25\,ms with 50\% of overlap, and FFT of 512 points in the 0-800\,Hz frequency band, since murmurs are rarely manifested in higher frequencies \cite{mcgee2018auscultation}. The choice of this specific frequency band was the only filtering in the frequency domain on PCGs. No further preprocessing was applied (e.g., spike removal algorithms) in order to avoid distorting murmur components. After passing through the Mel filter bank, a total of 32 values were obtained per frame.

%\begin{enumerate}
%    \item Each window was divided into frames of 25\,ms with an overlap of 50\% and windowed using a Hamming window; similar values were used in \cite{Zabihi16} and \cite{Rubin:2017}. Since the aim of the algorithm is to detect the murmur grading in the systolic phase, which is always longer than 25\,ms, this value was considered appropriate. Moreover, fundamental heart sounds are usually longer than 60\,ms, so this window size should be enough to capture murmur features. 
%    \item The magnitude of the discrete Fourier transform is calculated using 512 points in the 0-800\,Hz band, since murmurs are rarely manifested in higher frequencies\,\cite{mcgee2018auscultation}.
%    \item Each representation passes through a Mel filter bank to map onto the Mel scale, obtaining a total of 32 values per frame.
%    \item The Mel spectrogram is converted to decibels (dB).  
%\end{enumerate}

Thus, the 2-D representation of each 3\,s segment of heart sound recording is a 32$\times$239 matrix. In this study, the Mel spectrogram representation of the PCG was chosen, because it is commonly accepted by the audio processing community due to its similarity to human auditory perception. This is a reasonable choice, since the algorithm is expected to replicate human experts' decisions. Note that the choice of the Mel spectrum for PCG classification does not imply that the Mel spectrogram necessarily outperforms other 2-D representations of the PCG, such as wavelet transformation or Kalman-based spectro-temporal estimation~\cite{zhao2020kalman}. Figure~\ref{fig:data_examples} shows 1-D and 2-D representations of the PCGs for examples without murmur, with \textit{soft} murmur, and with \textit{loud} murmur. The 2-D representation is the input of the \ac{DNN}. 

For the sake of training the neural network, recording-level labels were generated. Note that the clinically significant labels are patient-level labels. By convention, we consider that each recording had the same label as the patient-level, except those recordings/locations where the annotator could not hear the murmurs. In that case, the labels were \textit{absent}. Then, the analysis windows of each recording inherited the recording-level label.

%In order to have recording-level murmur grading labels, by convention, we assumed that each recording had the same label as the patient-level. However, as it was already discussed, the recordings/locations where the annotator could not hear the murmurs were labeled as \textit{absent}. Note that some specific patients may have \textit{soft} murmurs but if these murmurs are audible on 3 recordings out of 4, the fourth recording will be labeled as \textit{absent}. So, despite having a single patient-level label, recording-level labels may be different within the same patient. For those patients without murmurs, all the recordings were labeled as \textit{absent}, and for those patients with \textit{loud} murmurs all the recordings were labeled as \textit{loud}.% So, it is possible that for a specific patient whose patient-level label was soft three recordings had label soft but one recording had label absent. xxx yyy zzz

%In addition, for training the neural network, all analysis windows of each recording inherited the recording-level label. 

\subsection{Neural Network Classifier}
\begin{figure}[tb]
\centering
\subfigure[Full architecture]{\includegraphics[scale=1]{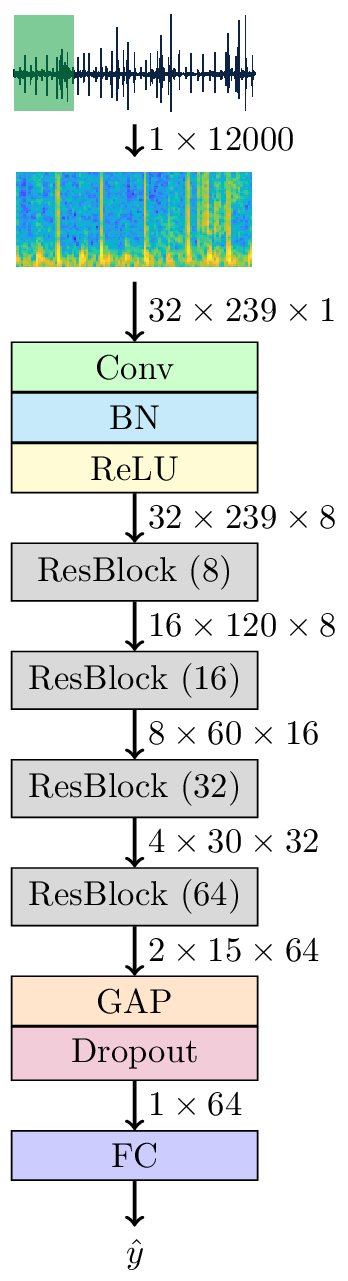}\label{fig:dnn_full}}
\subfigure[ResBlock]{\includegraphics[scale=1]{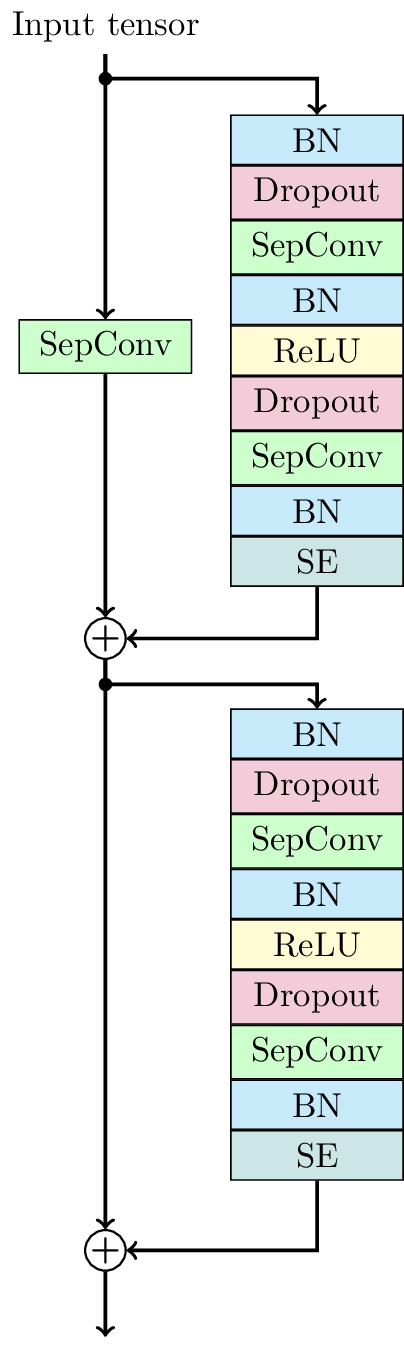}\label{fig:dnn_resblock}}
\caption{The full architecture of the residual neural network used in this study (panel a) and the architecture of each residual block (panel b).}
\label{fig:dnn}
\end{figure}

A fully convolutional residual neural network (ResNet) with channel-wise attention was used to classify the 2-D representation of each 3\,s window. The ResNet architecture relies on shortcut paths from and to layers at different stack positions, in order to diminish vanishing gradient phenomena during the training stage. The full architecture is shown in Figure~\ref{fig:dnn}, which is composed by the following layers:
\begin{itemize}
    \item \textbf{Convolutional layer (Conv):} The proposed architecture contains a single convolutional layer of order 4$\times$4 that generates 8 different representations of the input (feature maps). %Applies the convolution operation. The proposed architecture contains a single Conv layer of order $4\times4$ that generates 8 different representations of the input.
    \item \textbf{Batch normalization (BN):} Channel-wise BN was used as proposed in \cite{ioffe2015batch}, in order to speed up the training process, enhance generalization, and reduce the need for hyper-parameter tuning. After BN, a rectified linear unit (ReLU) function was applied to the outputs of the BN block in some cases (see Figure \ref{fig:dnn} for the BN blocks, which are followed by a ReLU). % These layers stabilize the distributions of the outputs between layers by applying a channel-wise normalization. This steps speed ups the training, improve generalization and reduces the need of fine hyper-parameter tuning. 
    % \item \textbf{Rectifier linear unit (ReLU)} % This layer applies the ReLU function to the input data.
    \item \textbf{Dropout:} The dropout rate was 20\% before the classification layer and 5\% in the residual blocks. \cite{srivastava2014dropout} recommended using $L_2$ regularization with dropout to avoid overfitting \cite{srivastava2014dropout}, so the rest of the layers were trained using a regularization term of $10^{-3}$.
    \item \textbf{Separable convolution (SepConv):} SepConv layers perform depth-wise and point-wise convolutions to generate $M$ representations of the inputs \cite{chollet2017xception}. Using SepConv layers, instead of traditional convolutional layers, reduces the number of the trainable parameters of the network, thus avoiding overfitting. In this architecture, the order of the depth-wise convolution was 1 and the order of the point-wise convolution was $L\times L$. In Figure~\ref{fig:dnn}, $M$ is shown for each residual block and $L$ is 1 in every shortcut, while $L=4$ is adopted to transform the data. When changing the dimension of the output for the first time (when SepConv is present in the shortcut in Figure~\ref{fig:dnn}), a stride of 2 was used in order to reduce the dimension.
    \item \textbf{Squeeze-and-excitation (SE):} These layers are channel-wise attention mechanism layers that adaptively recalibrate feature responses \cite{hu2018squeeze}. This is done in two steps: \textit{squeeze}, where the global information of each channel is embedded using global average pooling, and \textit{excitation}. For excitation, first two fully-connected layers are applied. The first one is composed of $M/r$ units, $r$ being the reduction ratio, and a ReLU activation function. The second fully-connected layer is composed of $M$ units with a sigmoid activation function. Finally, the output of the SE block is obtained by applying channel-wise multiplication between the outputs of the second fully-connected layer and the input feature map. The reduction ratio was 8 ($r=8$) in all layers. Lower values of $r$ did not improve the performance during the first experiments within the training set, but increased the number of trainable parameters.
    \item \textbf{Global average pooling (GAP):} The mean value for each channel was computed as proposed in the original ResNet \cite{he2016deep}, to obtain a feature vector of 64.
    \item \textbf{Fully-connected layer (FC):} A fully-connected layer with three output neurons and softmax activation to perform the classification task.
\end{itemize}

To train the neural network, the Adam optimizer was used with a learning rate of $0.001$, $\beta_1=0.9$, $\beta_2=0.999$ and $\epsilon=10^{-7}$, as suggested by \cite{kingma2014adam}; where $\beta_1$ and $\beta_2$ denote the exponential decay rates for the first and second moment estimates, respectively, and $\epsilon$ is a small constant for numerical stability. The number of epochs was fixed to 15 and the loss function was the standard categorical cross-entropy. The training batch size was 128. In order to address the issue of class-imbalance, the number of instances per class within each mini-batch was equalized. At each epoch, the majority class (\textit{absent}) was randomly under-sampled without replacement and the minority classes (\textit{soft} and \textit{loud}) were randomly over-sampled until reaching a specific size, which is defined by the batch size and the number of iterations. %In order to avoid class imbalance, the number of instances within each mini-batch was equalized by under-sampling the majority classes.

Every hyper-parameter of the classifier, including filter order, number of filters, and dropout ratios, was tuned using 10-fold cross-validation and a grid-search on the training set to maximize the performance metrics explained in Section \ref{sec:evaluation_criteria}. The search ranges were $\{2,3,\ldots,6\}$ for $L$, $\{4,8,\ldots,24\}$ for the initial $M$, $\{0.05, 0.1,\ldots,0.4\}$ for the dropout ratios, $\{2,4,8\}$ for $r$, $\{32,64,\ldots,256\}$ for the batch size, and $\{5, 15, 25, 50, 75, 100\}$ for the number of epochs.

\subsection{Ensemble Learning} \label{sec:ensemble_learning}
Ensemble learning is one of the oldest and most powerful techniques in supervised machine learning \cite{rokach2010ensemble,hansen1990neural}. In this case, ensemble learning consists of training multiple independent DNNs. Training a neural network model using the same training data may lead to different results due to different random initialization, which end in different local minima after the training process~\cite{allen2020towards}. Nevertheless, this characteristic can be used to boost the performance of the final classifier. 

In this study, we used a 15-fold patient-wise cross-validation committee to train the neural networks for each training subset. We generated 15 replicates of the training data and removed 1/15 disjoint subsets of the data from each replica. By doing this, we would have 15 subsets of the training data, and each subset was slightly different from another subset. Each of these subsets was used to train each of the 15 models. At the end, a total of 15 models were trained using the same architecture, but with different random seeds to initialize the trainable parameters of the model and slightly different training data ($14/15\approx 93\%$ of the training data are used in each set). 

When applying the method to validation or test data, the softmax outputs of 15 neural networks are computed for each window, i.e. $p_i=\{p_{i,\mathrm{absent}},p_{i,\mathrm{soft}},p_{i,\mathrm{loud}}\}$ for $i={1,2,\ldots,15}$. In order to make a decision about the window of 3\,s, the arithmetic mean is computed over fifteen models to obtain three values, the \textit{likelihoods} associated to each of the classes: %The window is classified as the class with the maximum likelihood. 
% \begin{equation} \label{eq:pw_absent}
%     p_{w,\mathrm{absent}}=\frac{1}{N}\sum\limits_{i=1}^{N} p_{i,\mathrm{absent}}, 
% \end{equation}
% \begin{equation} \label{eq:pw_soft}
%     p_{w,\mathrm{soft}}=\frac{1}{N}\sum\limits_{i=1}^{N} p_{i,\mathrm{soft}}, 
% \end{equation}
% \begin{equation} \label{eq:pw_loud}
%     p_{w,\mathrm{loud}}=\frac{1}{N}\sum\limits_{i=1}^{N} p_{i,\mathrm{loud}}. 
% \end{equation}
\begin{equation} \label{eq:pw}
     p_{w,c}=\frac{1}{N}\sum\limits_{i=1}^{N} p_{i,c}, 
\end{equation}
where $N=15$ is the number of classifiers and $c \in \{absent, soft, loud\}$.

\subsection{Final Decision Rule}

The analysis of the murmur's severity and its corresponding grade are achieved by a joint analysis of several recordings from different auscultation locations. As a result, in our signal processing pipeline, it is necessary to merge the class continuous output (calculated using Equation\,\ref{eq:pw}) obtained for each 3\,s window to make a decision about the recording first, and then make an overall decision about the patient.

For each recording containing $j=1,\ldots,N_w$ 3\,s windows, the three $p_w^{(j)}$ values for each window were first obtained through Equation \eqref{eq:pw}: $p_{w,absent}^{(j)}$, $p_{w,soft}^{(j)}$ and $p_{w,loud}^{(j)}$. Then, the arithmetic means of each $p_w$ over $N_w$ windows were computed, and the category associated with the maximum likelihood was assigned to the entire recording. To assign a class to a patient, the most severe grade level detected among all recording locations of the same patient was assigned to the patient, with an exception: in the case of patients for whom only some of the auscultation locations were available, even when the most severe grade was \textit{loud}, we automatically assigned the \textit{soft} label in order to be consistent with the annotation criteria described in Section~\ref{par:database}. This labeling strategy is summarized in Algorithm \ref{alg1}. Note that the recording-level results are intermediate results of the proposed algorithm, but the overall (clinical outcome) are the patient-level results.
 
\begin{algorithm}[h]
\caption{Final Decision Rule for Murmur Grading}\label{alg1}
\begin{algorithmic}
%\Require $n \geq 0$
%\Ensure $y = x^n$
%\While{$N \neq 0$}
\If {recordings from all locations are available}
   
    \State {patient-level label is \textit{the most severe label of recordings}} 
\ElsIf{all recordings are classified as absent}
    \State {patient-level label is \textit{absent}}
    \Else
    \State {patient-level label is  \textit{soft}}
     \Comment{See Section~\ref{par:database}}
\EndIf
%\EndWhile
\end{algorithmic}
\end{algorithm}

\subsection{Estimation of the uncertainty}

The ensemble of CNNs also provided information about the \textit{uncertainty} of the classifier's decision. We computed the standard deviation of $p_i$ for three classes in the window level, to obtain $\gamma_i^{\mathrm{(absent)}}$, $\gamma_i^{\mathrm{(soft)}}$ and $\gamma_i^{\mathrm{(loud)}}$. The mean among these three values was computed, resulting in the window-level uncertainty measure $\gamma_i$. This uncertainty measure can be easily extended to the recording-level, by computing the mean of all $\gamma_i$ values within the same recording ($\gamma_r$). Finally, patient-level uncertainty, $\gamma_p$, was obtained by computing the overall mean among all $\gamma_r$ values for the same patient.

\subsection{Evaluation Criteria} \label{sec:evaluation_criteria}

The models were first evaluated using patient-wise stratified 10-fold cross-validation in the training subset. The computed performance metrics were the sensitivity for each class, the unweighted (arithmetic) mean of sensitivities (UMS), the F1-score per class, and the arithmetic mean of F1-scores. %Every metric was computed in recording-level and patient-level.

Then, the proposed algorithm was validated using the test set of the George B. Moody Physionet Challenge 2022\,\cite{reyna2022heart}. During the training process a total of 150 models were trained (10-fold cross-validation and 15 models at each iteration; Section \ref{sec:ensemble_learning}), which were used to detect murmur's severity grading in the test data. A single run was performed on the test subset.

\subsection{Adaptation to normal/abnormal PCG classification}

The same method was also tested for another different task: normal/abnormal PCG classification using data from the 2016 Physionet/Computing in Cardiology Challenge \cite{clifford2016classification,liu2016open}. The task consisted of detecting abnormalities in the PCG recordings. So the three neurons in the FC layer of the proposed model (see Figure\,\ref{fig:dnn}) were replaced by two neurons, and the rest were untouched. Again, using windows of 3\,s, an ensemble of neural networks made decisions; then the mean values among the recordings were calculated to make the final decision. The method was evaluated in terms of the challenge score: modified versions of sensitivity and specificity were calculated first, and the final score was the mean value among both\,\cite{liu2016open,clifford2016classification,Clifford17}. Note that the \textit{unsure} class was not considered, but the challenge metric could be computed anyway.

\section{Results}
\label{par:results}

\subsection{The Performance of the Proposed Model}
%The overall performances using 10-fold cross-validation for patient-level murmur gradings were 86.3\% and 81.6\% in terms of the UMS and average F1-scores. Due to our final decision policy for murmur grading (see Algorithm~\ref{alg1}), the overall patient-level performances were significantly better than overall recording-level performances (i.e., 79.6\% and 77.8\% in terms of the UMS and average F1-scores).
The overall performances using 10-fold cross-validation
for patient-level murmur gradings were 86.3\% and 81.6\% in
terms of the UMS and average F1-scores. Due to our final
decision policy for murmur grading (see Algorithm~\ref{alg1}), the
overall patient-level performances were significantly better
than overall recording-level performances (i.e., 79.6\% and
77.8\% in terms of the UMS and average F1-scores), because of a better performance when detecting \textit{soft} and \textit{loud} murmurs. However, the Se and F1-score for \textit{absent} class were better in recording-level. Note that the proposed algorithm requires detecting \textit{absent} in all the recordings from the same patient in order to assign the label \textit{absent} to the patient. Thus, the patient-level Se for \textit{absent} class was poorer than the recording-level one (90.7\% vs 95.0\%), and despite resulting in a slightly better patient-level positive predictive value (96.8\% vs 96.5\%), the patient-level F1-score was for \textit{absent} class was poorer than the recording-level one (93.6\% vs 95.7\%). Again, it is worth mentioning that the clinically relevant labels are patient-level labels.

\begin{figure}[tb]
    \centering
    \includegraphics[width=.8\linewidth]{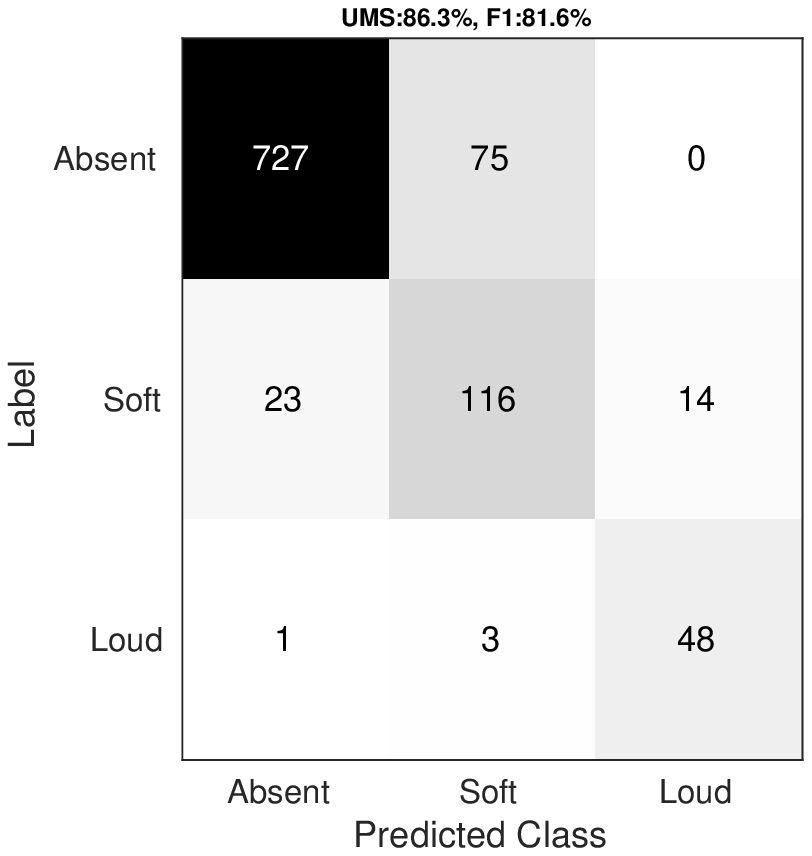}
    \caption{Confusion matrix for classification using 10-fold cross-validation.}
    \label{fig:my_label}
    \label{fig:CM}
\end{figure}

Figure~\ref{fig:CM} shows the detailed patient-level performance in the form of the overall confusion matrix. The resulting patient-level sensitivities (F1-scores) were 90.7\% (93.6\%), 75.8\% (66.8\%), and 92.3\% (84.2\%), respectively for detecting the \textit{absent}, \textit{soft}, and \textit{loud} classes. As expected, \textit{soft} murmurs were the most challenging class to detect correctly. There are 98 (=75+23) misclassified cases between \textit{absent} and \textit{soft} classes. The misclassified cases between \textit{soft} and \textit{loud} classes are 17 (=14+3). However, only one patient was misclassified between \textit{absent} and \textit{loud} classes. The PCG recordings of this subject were identified to be noisy. Moreover, the uncertainty of the network was relatively high (cf. Section \ref{sec:methods_uncertainty}).

To put it in perspective, the proposed method was also cross-validated on the publicly available training set of the 2016 PhysioNet Challenge\,\cite{liu2016open} using 10-fold cross-validation and the resulting challenge score was around 90\%. This performance is similar to those obtained by the top two performing algorithms in the challenge in their validation sets within the training set \cite{potes2016ensemble,Zabihi16}.

\subsection{The Impact of the Analysis Window}

The recordings of the training set contained a median of seven windows of 3\,s, which were used for algorithmic decision making. We observed that the recording-level performance was correlated with the number of analysis windows, and the obtained performance was better in longer recordings as more windows were analyzed. For instance, the UMS and mean F1-score were 78.1\% and 76.4\% for those recordings with less or equal to seven windows. On the other hand, UMS and mean F1-score were 80.8\% and 79.3\% for those recordings with more or equal to seven windows. Also, considering only the recordings with $\geq7$ windows and analyzing the first three to five windows, the recording level UMS  and F1-scores decreased approximately by 1 to 3 percentage points.  

%to 78.3\% (76.0\%) and 79.8\% (78.2\%), respectively.

When we made the analysis with longer windows of 5\,s (resulting in fewer windows per recording) or smaller windows of length 1\,s, the cross-validated performances decreased, demonstrating the fact that the window size should be long enough to capture information about consecutive beats and the murmurs. A window size of 3\,s was found to be a good compromise between the number of windows analyzed per recording and the provided information about the beats and murmurs. The same window size was reported in a recent state-of-the-art algorithm for PCG processing \cite{Xiao20}.

\subsection{The Effect of Neural Networks Ensemble} \label{sec:methods_ensembleDNN}
\begin{figure}[tb]
    \centering
    \includegraphics[width=0.8\linewidth]{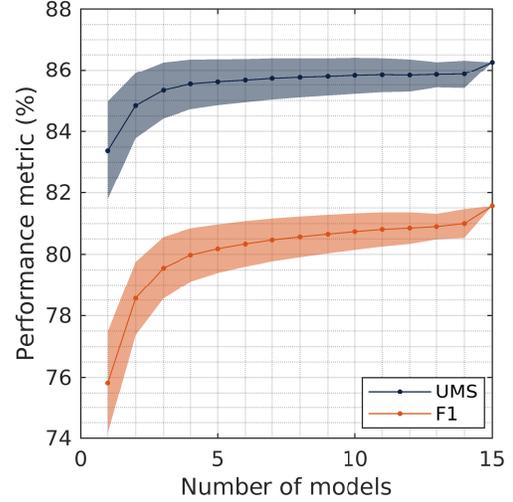}
    \caption{Patient-level unweighted mean of sensitivities (UMS) and mean F1-scores in function of the number of the models used in the ensemble.}
    \label{fig:ensemble_results}
\end{figure}

The proposed method was based on an ensemble of 15 neural networks, and the combination of all 15 networks showed better performance than a single network, in average. Figure~\ref{fig:ensemble_results} shows the cross-validated patient-level UMS and the mean F1-scores as a function of the number of models in the ensemble. Having 15 trained models at hand, by choosing $n_m$ models ($1<n_m<15$) from them, all possible $\displaystyle{15 \choose n_m}$ combinations were considered. For each combination, the performance metrics were computed. The mean for each $n_m$ value is shown in Figure~\ref{fig:ensemble_results} with dots, and the shadowed area represents $\pm$standard deviation. A single model achieved the UMS and the average F1-score of 83.4\% and 75.8\%, respectively. Using 15 models in the ensemble, the UMS and the average F1-score improved by more than 2.8 and 5.7 percentage points. Individual F1-scores for each class improved from 89.7\% to 93.6\%, from 57.3\% to 66.9\%, and from 80.4\% to 84.2\% for \textit{absent}, \textit{soft}, and \textit{loud} murmurs, respectively. %The median for each $n_m$ value is shown in Figure~\ref{fig:ensemble_results}. A single model achieved the UMS and the average F1-score of 83.9\% and 75.9\%, respectively. Using 15 models in the ensemble, the UMS and the average F1-score improved by more than 2.5 and almost 6 percentage points. Individual F1-scores for each class improved from 90.1\% to 93.6\%, from 56.8\% to 66.9\%, and from 80.7\% to 84.2\% for \textit{absent}, \textit{soft}, and \textit{loud} murmurs, respectively. 

The performance of the model was further cross-validated by increasing the number of models above 15 up to 20. However, while the UMS increased by 0.2 percentage points for 20 models, the F1-score decreased by 0.9 percentage points.

\subsection{Uncertainty analysis}
\label{sec:methods_uncertainty}
For $\gamma_i$, $\gamma_r$ and $\gamma_p$, the uncertainty was (statistically) significantly higher for the miss-classified cases in the 10-fold cross-validation scheme. For instance, the median (IQR) $\gamma_p$ was 0.11 (0.08--0.14) for correctly classified patients and 0.15 (0.13--0.18) for misclassified patients ($p<0.001$). For the patient that was misclassified between \textit{absent} and \textit{loud} classes, the uncertainty was relatively high, with $\gamma_p=0.23$. The proposed approach is important for prescreening applications, as it provides information not only about the grading of the murmur, but also about the decision confidence made by the artificial intelligent agent. The clinical staff may use these algorithmic confidences, in their final decision-making. For instance, a higher performance can be obtained for cases with low uncertainty values and an expert could review the uncertain cases.

\begin{figure}[tb]
    \centering
    \includegraphics[width=.8\linewidth]{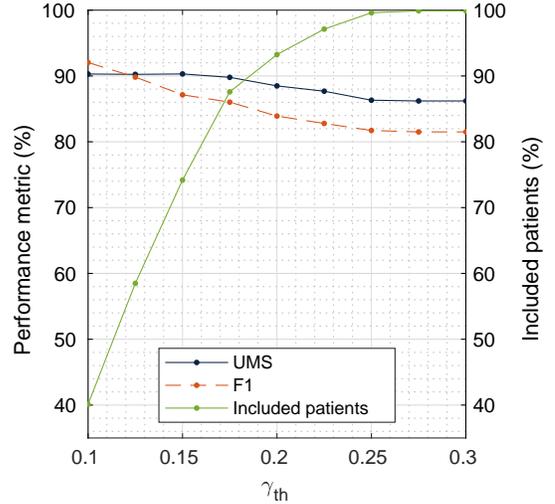}
    \caption{Performance metrics and percentage of included patients when considering only those patients with $\gamma_p<\gamma_{th}$}
    \label{fig:uncertainty}
\end{figure}

To further demonstrate the effect of the uncertainty measure, the uncertainty threshold $\gamma_{th}$ was fixed and the performance metrics were calculated only using those patients with $\gamma_p\leq\gamma_{th}$. The results are shown in Figure \ref{fig:uncertainty}. %shows the obtained results. 
It can be observed that for lower values of $\gamma_{th}$ the performance metrics are better, but the percentage of included patients is lower. For instance, discarding less than 7\% of the patients boosted the UMS and average F1-scores by approximately 2 percentage points. 

\subsection{Results on the hidden test data}

\begin{figure}
    \centering
    \includegraphics[width=.8\linewidth]{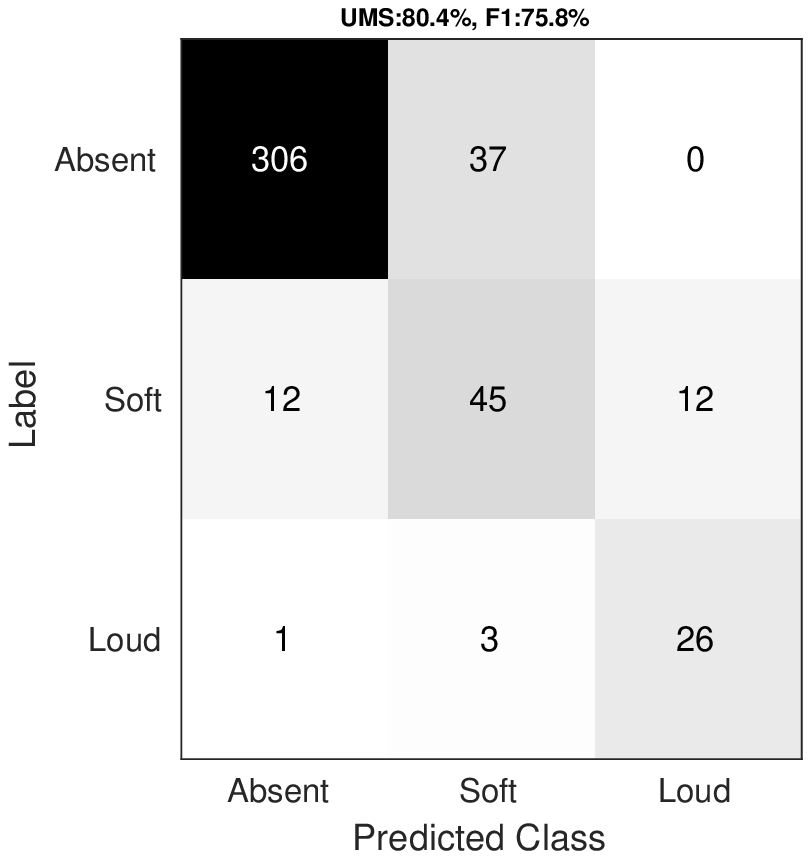}
    \caption{Results on the test set}
    \label{fig:confmat_test}
\end{figure}

The obtained confusion matrix for the test set is shown in Figure\,\ref{fig:confmat_test}. Overall UMS and F1-scores were 80.4\% and 75.8\%, both dropped $\approx5$ percentage points compared to the 10-fold cross-validation performed with the training data. Again, only a single patient was misclassified between \textit{absent} and \textit{loud} classes.

Longer recordings showed better recording-level performance, as in the training set. For instance, those recordings with more than 7 windows showed a UMS and F1-score of 77.7\% and 75.4\%, respectively, while those recordings with less than 7 windows showed a UMS and F1-score of 74.2\% and 73.3\%.

Finally, the estimated uncertainty showed a median (IQR) of 0.12 (0.10-0.15) for correctly classified patients and 0.16 (0.14-0.19) for those patients that were not correctly classified. Although the numbers changed slightly, the uncertainty measurement still showed statistically significant differences between correctly classified patients and misclassified patients ($p<0.001$).

\subsection{Comparison with other algorithms}

Although we cannot compare our methods with other state-of-the-art algorithms due to mismatches in their objectives, we can still adapt the output of our algorithm to calculate the performance of the binary classification (\textit{murmurs absent} vs. \textit{present}) for comparison. If we group \textit{soft} and \textit{loud} classes into \textit{present murmurs}, then the cross-validated sensitivity and specificity of murmur detection would be 88.3\% and 90.7\%, respectively (see Figure \ref{fig:CM}). For the test set, the sensitivity and specificity would be 86.9\% and 89.2\%, respectively (see Figure\,\ref{fig:confmat_test}). These numbers are in line with the performance of the state-of-the-art algorithms. However, one should be aware that the performance of the murmur detection algorithms strongly depend on the training and test datasets; more specifically, the performance of the algorithms depend on the amount of \textit{soft} murmurs present in the dataset.

\section{Discussion}
\label{par:discussion}

\subsection{Murmur Grading}
Murmur grading is one of the fundamental steps toward the comprehensive characterization of murmurs. The most common method to characterize the grading is the Levine scale~\cite{Levine:1933}; but due to its complexity, alternative murmur grading scales have been explored by the medical community~\cite{rishniw2018murmur}. Within the biomedical engineering and machine learning communities, many efforts have been made to detect the presence or absence of murmurs\,\cite{chorba2021deep, alam2018murmur}; but automatic methods of murmur grading have not been proposed yet. This study presented a fully automatic algorithm for murmur grading.

%The scale to characterize the murmur was based on Levine scale; although it was slightly modified for the following reasons. 

Murmur grading, per se, has clinical significance, since louder murmurs are associated with different cardiac pathologies. Softer murmurs may be innocent or pathological, and they may manifest for the first time during the early stages of many cardiac diseases. In fact, a recent study showed that males with \textit{soft} murmurs had an increased risk of aortic valve replacement \cite{bodegard2012low}. 

The dataset used in this work was recorded during a public prescreening campaign in a rural area, and detecting \textit{soft} murmurs in such scenarios may lead to earlier detection of many cardiac disorders, which may improve the quality of life of the patients and reduce costs. However, \textit{soft} murmurs were the most challenging to detect correctly, which has been observed by another recent study too. For instance, Chorba et al.\,\cite{chorba2021deep} reported that the sensitivity to detect the presence of murmurs improved from $76.3\%$ to $90.0\%$ when discarding grade I murmurs.

\subsection{Segmentation vs. No-segmentation}
Many algorithms use a segmentation step in order to extract information from the PCG, including algorithms based on DNNs \cite{chen2021deep}, while other research have not applied any segmentation as an intermediate step \cite{Zabihi16}. In this study, no segmentation algorithm was used, due to two key challenges. Firstly, the recordings were captured in a mobile setting with varying conditions. Factors such as background noise complicates the automatic segmentation task. Secondly, automatic segmentation algorithms usually perform worse when murmurs are present, and their error may propagate into the main algorithm and reduce the performance of any classification algorithm. 

\subsection{The Choice of DNN}
Separable convolutions reduced the number of trainable parameters ($\approx$33,000 in the proposed architecture) and possibly improve the generalization capability. Using regular convolutions increased the number of trainable parameters ($\approx$312,000) and led to a similar cross-validated F1-score (81.6\% vs. 82.1\%), but the UMS decreased from 86.3\% to 82.7\%. The highest decrease was observed in the sensitivity of \textit{soft} murmurs, from 75.8\% to 60.8\%. Adding SE blocks also improved the results, the same architecture without these blocks lead to a cross-validated UMS and F1 scores of 85.3\% and 80.4\%, 1 percentage point below the best cross-validated scores.

We also tested other algorithms and architectures that achieved good results for binary PCG classification over other datasets and tasks (e.g., normal vs abnormal classification), including solutions based on 1-D CNNs \cite{Xiao20}, VGG-like CNNs \cite{maknickas2017recognition}, and solutions based on hand-crafted features without previous segmentation \cite{Zabihi16}. However, in all the cases we obtained inferior results.

The proposed algorithm was able to estimate the uncertainty of the class labels. It is practically very important to report algorithmic diagnosis together with their confidence/uncertainty. This helps experts in confirming or rejecting algorithmic outcomes and in their decisions regarding the border cases that require further investigation by medical experts.

\subsection{Study Limitations and Future Work}
This study has four main limitations, which are associated with the utilized dataset. The first limitation is that a single expert annotated all the data. Murmur grading annotations are subjective, and having more annotators would reduce the bias. Secondly, the patients used in this study are limited to a specific population, multi-center studies are needed to further confirm the obtained results. Thirdly, the murmur grading was divided into three different groups; thus, a single grading mechanism was tested. Future studies should also consider other grading scales to analyze the reliability of the algorithms. Finally, the algorithm was tested using systolic murmurs only. Although it may be reasonable to infer that the algorithm should work well during diastolic murmurs, further studies are needed to confirm this in the future.

Another possible future line of research is the full characterization of murmurs, which includes the automatic estimation of other characteristics such as timing, shape, pitch, or quality in addition to murmur grade. Since the dataset used in this study includes those ground truth labels \cite{oliveira2021circor}, the development of automated algorithms is feasible in the future.

\section{Conclusion}
\label{par:conclusions}
This paper presented a novel algorithm for multi-class murmur detection and murmur grading, based on heart sounds, in a population of mostly pediatric patients. In the future, implementing the proposed algorithm on edge devices would support clinicians for pre-screening purposes, not only giving feedback about the grading but also measuring the uncertainty about the decision made by the algorithm.

\bibliographystyle{IEEEtran.bst}
\bibliography{example}

\end{document}